\documentclass[twocolumn,aps,pr,superscriptaddress,preprintnumbers,nofootinbib,10pt]{revtex4-2}
\usepackage{amsmath}
\usepackage{amssymb}
\usepackage[dvipdf,dvips]{graphicx}
\usepackage{color}
\usepackage{hyperref}
\usepackage{url}
\usepackage{slashed}
\usepackage{subfigure}
\usepackage[usenames,dvipsnames]{xcolor}
\usepackage{amsmath}
\usepackage{amsfonts}
\usepackage{float} 
\usepackage{amssymb}
\usepackage{epsfig}
\usepackage{graphics}
\usepackage{euscript}
\usepackage{slashed}
\usepackage{epstopdf}
\usepackage[utf8]{inputenc}
\allowdisplaybreaks
\usepackage[normalem]{ulem}
\usepackage{pifont}
\usepackage{dsfont}
\usepackage{MnSymbol}
\usepackage{verbatim}
\usepackage{graphicx}
\usepackage{latexsym}
\usepackage{tikz-feynman}

\begin{document}

\title{Entangled coherent states and violations of Bell-CHSH inequalities}
	
\author{Philipe De Fabritiis} \email{pdf321@cbpf.br} \affiliation{CBPF $-$ Centro Brasileiro de Pesquisas Físicas, Rua Dr. Xavier	Sigaud 150, 22290-180, Rio de Janeiro, Brazil}

\author{Fillipe M. Guedes} \email{fillipe.guedes@ufv.br} \affiliation{UERJ $–$ Universidade do Estado do Rio de Janeiro, Rua São Francisco Xavier 524, 20550-013, Maracanã, Rio de Janeiro, Brazil}

\author{Giovani Peruzzo} \email{gperuzzofisica@gmail.com} \affiliation{UFF $-$ Instituto de F\'{i}sica, Universidade Federal Fluminense, Campus da Praia Vermelha, Av. Litor\^{a}nea s/n, 24210-346, Niter\'{o}i, Brazil}
	
\author{Silvio P. Sorella} \email{silvio.sorella@gmail.com} \affiliation{UERJ $–$ Universidade do Estado do Rio de Janeiro, Rua São Francisco Xavier 524, 20550-013, Maracanã, Rio de Janeiro, Brazil}
	
\begin{abstract}
Three classes of entangled coherent states are employed to study the Bell-CHSH inequality. By using pseudospin operators in infinite dimensional Hilbert spaces, four dichotomic operators $(A,A',B,B')$ entering the inequality are constructed. For each class of coherent states, we compute the correlator $\langle \psi \vert A B + A' B + A B' - A' B' \vert \psi \rangle$, analyzing the set of parameters that leads to a Bell-CHSH inequality violation and, particularly, to the saturation of Tsirelson's bound.

\end{abstract}

\maketitle	
	
\section{Introduction}\label{Introduction}
\vspace{-0.1cm}

The advent of Quantum Mechanics has changed our way of seeing the world, showing that Nature can be subtle and does not always work in the way our intuition suggests. A striking example of a non-intuitive feature present in Nature is the phenomenon of entanglement~\cite{EPR}, that is, the existence of  states of a composite system that cannot be written as a product of the states of individual subsystems~\cite{Horodecki09}. 
An entangled system has quantum correlations between its constituents that cannot be accounted for by any local realistic theory, as shown by Bell in 1964 through the formulation of his renowned inequality~\cite{Bell64,Bell66}. There is a particularly interesting and popular version of Bell's inequality that is more suitable for experimental verifications, the so-called Clauser-Horne-Shimony-Holt (CHSH) inequality~\cite{CHSH69,CHSH74,CHSH78}. The CHSH inequality is violated by entangled states in Quantum Mechanics~\cite{ExpCHSH1, ExpCHSH2, ExpCHSH3, ExpCHSH4, ExpCHSH5, ExpCHSH6, ExpCHSH7, ExpCHSH8, ExpCHSH9}, with its maximum violation being given by Tsirelson's bound~\cite{Tsirelson80, Tsirelson87}: $2 \sqrt{2}$.
The existence of entanglement can be considered the deepest departure from classical physics~\cite{Werner01},  having far-reaching consequences, both from the theoretical and technological sides~\cite{Yu21}. 

Among the large set of quantum states which exists in a given Hilbert space, the so-called coherent states~\cite{Schrodinger26,Glauber63, Sudarshan63, Klauder85} exhibit properties which, in the large occupation number, allow us to regard them as the most classical objects that one can devise in  a quantum system. Yet, systems described by these states can exhibit entanglement, leading to relevant developments in many areas such as quantum information and quantum optics~\cite{SandersReview}. 
There is a great interest in superpositions of coherent states, a subject that first appeared in~\cite{Milburn85,Milburn86}, and was analyzed in more detail later in~\cite{Yurke86,Yurke87}. The production of such states in the case of a single mode of the electromagnetic field was studied in~\cite{Glancy08}, and their main properties and extensions in~\cite{Titulaer65,Birula68,Stoler71,Barut71,Arecchi72,Gilmore72,Perelomov72,Perelomov86}. For a review on this subject, see~\cite{Buzek95}. 

Entangled coherent states first appeared in 1967~\cite{Aharonov67} and had to wait almost twenty years for their next appearance, in~\cite{Yurke86}. The so-called pair coherent state~\cite{Agarwal86,Agarwal88,Agarwal91} appeared in the same year, being a special case of the already known state considered in~\cite{Barut71}. Although  these states were somehow present in the literature since 1967, the first time entangled coherent states were directly studied was in~\cite{Mecozzi87,Tombesi87}, where the authors generalized to multimode coherent states what was done in~\cite{Milburn85,Milburn86,Yurke86}. The term {\it entangled coherent state} was introduced only in 1992~\cite{Sanders92}, in a paper that analyzed the Bell-CHSH inequality violation for such states and also how to produce them. In that work, the few-photon limit was considered, but later on, it was shown that entangled coherent states violate the Bell inequality in the large photon number limit as well~\cite{Mann95}.

These states can be generalized to superpositions of multimode coherent states~\cite{Jex95,Zheng98,Wang01b}. There are important examples of multipartite states such as the Greenberger–Horne–Zeilinger (GHZ) and W-types of states~\cite{Jeong06,Li09} as well as the cluster states~\cite{Munhoz08,Becerra08,Wang08}.
One can also consider more general entangled coherent states upon considering abstract generalizations of coherent states~\cite{Arecchi72,Gilmore72,Perelomov72,Perelomov86}. It is worth underlining that entangled coherent states are currently employed in quantum teleportation~\cite{Bennett93,Wang01a,Johnson02,vanEnk01,Jeong01,Janszky02},  in quantum information processing~\cite{Munro01,Cochrane99,Oliveira00}, in quantum networks~\cite{vanLoock08,El Allati11} and  in quantum metrology~\cite{Ansari94a,Ansari94b,Gerry01,Gerry02,Joo11}.

The aim of this paper is to present a study of the Bell-CHSH inequality violation by considering a variety of entangled coherent states. We shall check out when  
\begin{align}\label{b1}
 \vert \langle \psi \vert \;	\mathcal{C}_{CHSH} \; \vert \psi \rangle \vert    > 2 \;, 
\end{align}
where $ \mathcal{C}_{CHSH} = (A +A')B + (A  - A' )B'$ is the Bell-CHSH operator and $\vert \psi \rangle$ stands for an entangled coherent state. Here, the dichotomic Bell operators are defined such that 
\begin{align}\label{b2}
	A^2 = B^2 = 1; \quad A^\dagger = A, \, B^\dagger = B; \quad \left[A,B\right] = 0,
\end{align}
with similar equations holding for the primed operators. These operators will be obtained by relying on the so-called pseudospin operators~\cite{psi1,psi2,psi3}, which reproduce the algebra of the Pauli matrices in the case of infinite dimensional Hilbert spaces needed for defining coherent states. 

This work is organized as follows. In Section \ref{ECS}  we give a brief account of entangled coherent states. Section~\ref{Bellop} is devoted to the construction of Bell operators in the infinite dimensional Hilbert space by means of the pseudospin operators. In Section~\ref{SecSymmCohe}, we perform a detailed investigation of the Bell-CHSH inequality violation for the so-called symmetric entangled coherent states. Section~\ref{SecAsymmCohe} deals with the asymmetric case. The Schr\"odinger cat states are studied in Section~\ref{SecCatCohe}. Finally, in Section~\ref{Conclusions}, we state our conclusions.

\section{Entangled Coherent States}\label{ECS}

Bosonic coherent states can be defined as the annihilation operator eigenstates and can be represented in the number state basis as
\begin{align}\label{key}
	\vert \alpha \rangle = e^{-\frac{\alpha^2}{2}} \sum_{n=0}^{\infty} \frac{\alpha^n}{\sqrt{n!}} \, \vert n \rangle \;,
\end{align}
where $\vert n \rangle = \frac{(a^\dagger)^n}{\sqrt{n!}} \vert 0 \rangle$ are the Fock basis states, and we have by construction $a \vert \alpha \rangle = \alpha \vert \alpha \rangle$. Here, $\alpha$ is taken as a real number. An equivalent formulation can be given by employing the unitary displacement operator ${\cal D}(\alpha)$: 
\begin{align}
&{\cal D}(\alpha) =  e^{(\alpha a^\dagger - \alpha a)} = e^{\frac{\alpha^2}{2}} e^{\alpha a^\dagger} e^{-\alpha a}, \nonumber \\
&{\cal D}^\dagger(\alpha) {\cal D}(\alpha)  = {\cal D}(\alpha) {\cal D}^{\dagger}(\alpha) = 1\;, \nonumber \\
&{\cal D}^\dagger(\alpha) \, a \, {\cal D}(\alpha) = a + \alpha  \label{displ}
\end{align}
The coherent state $\vert \alpha \rangle$ is obtained by acting on the vacuum state with the operator ${\cal D}(\alpha)$, namely,
\begin{equation} 
\vert \alpha \rangle = {\cal D}(\alpha) \vert 0 \rangle.  \label{Dop}
\end{equation}

Entangled coherent states can be built by means of the superposition of multimode coherent states and have a huge number of applications, as highlighted in the Introduction.
For instance, in optics, there is much interest in the use of two-mode maximally entangled number states, the so-called $N00N$ states
\begin{align}\label{key}
	\vert \psi \rangle_{ N00N} = \frac{1}{\sqrt{2}} \left[\vert N \rangle_a \vert 0 \rangle_b + e^{i \phi} \vert 0 \rangle_a \vert N \rangle_b\right],
\end{align} 
that can find applications in quantum metrology, quantum sensing, and quantum interferometric photolithography~\cite{Dowling08,Boto00}. In~\cite{Wildfeuer07}, the authors investigated possible non-local correlation experiments using $N00N$ states with a relative phase $\phi = \pi$, showing the violation of some Bell-type inequalities for a total number of photons $N$\footnote{The Bell-CHSH inequality for $N=1$ has been studied in~\cite{Banaszek99}.}. In~\cite{Gerry09}, the authors extended the analysis done in~\cite{Wildfeuer07} considering maximally entangled coherent states
\begin{align}\label{key}
	\vert \psi_\alpha \rangle = N_\alpha e^{-\frac{\vert \alpha \vert^2}{2}} \sum_{n=0}^{\infty} \frac{\alpha^n}{\sqrt{n!}} \left[\vert n \rangle_a \vert 0 \rangle_b + e^{i \left(\phi + n \theta\right)} \vert 0 \rangle_a \vert n \rangle_b\right],
\end{align}
where $N_\alpha = \frac{1}{\sqrt{2}} \left(1 + \cos\phi \, e^{-\vert\alpha\vert^2}\right)^{-\frac{1}{2}}\!\!$. These states can be considered as superpositions of $N00N$ states, and can be useful in the context of interferometry~\cite{Gerry00,Benmoussa02}. For the cases considered by them, it was found a greater degree of violation for the Bell-type inequalities considered, in comparison with the $N00N$ states results~\cite{Wildfeuer07}.

More elaborated states can be obtained by considering the symmetric and asymmetric entangled coherent states, a slight generalization of the cases considered in~\cite{Gerry09,Park15}. Here we define the symmetric case as
\begin{align}\label{Symc}
	\vert \psi \rangle_S = N_S \left[ \vert \alpha \rangle_a \vert \beta \rangle_b + e^{i \phi} \vert \beta \rangle_a \vert \alpha \rangle_b \right],
\end{align}
and the asymmetric case as
\begin{align}\label{AsymC}
\vert \psi \rangle_A = N_A \left[ \vert \alpha \rangle_a \vert \beta \rangle_b + e^{i \phi} \vert -\alpha \rangle_a \vert -\beta \rangle_b \right],
\end{align}
where $(N_S, N_A)$ stand for normalization factors which will be given below and  $\phi$ is a relative phase. 
Finally, we shall also consider entangled coherent states built out from Schr\"odinger cat states~\cite{Dodonov74}, {\it i.e.},
\begin{align}\label{Catc}
	\vert \psi \rangle_\pm = C_\pm \left[\vert \alpha \rangle_\pm \vert \beta \rangle_\pm + e^{i \phi}  \vert \beta \rangle_\pm \vert \alpha \rangle_\pm\right],
\end{align}
where $C_\pm$ is a normalization factor and we defined 
\begin{align}\label{Schcat}
	\vert \alpha \rangle_\pm = N_\pm \left[ \vert \alpha \rangle \pm \vert -\alpha \rangle  \right].
\end{align}

In what follows, we shall present a detailed investigation of the Bell-CHSH inequality violation, Eq.~\eqref{b1}, for both symmetric and asymmetric entangled coherent states, Eqs.~\eqref{Symc},\eqref{AsymC}, as well as for the cat states~\eqref{Catc}. 
We would like to emphasize that, although we used the specific states cited above to illustrate our method, we could apply it in principle to any entangled coherent state.

\vspace{-0.1cm} 
\section{Bell's operators construction}\label{Bellop}
 
The first goal to introduce the Bell-CHSH correlator, 
 \begin{align}\label{key}
	\mathcal{C}_{CHSH} = A B + A' B + A B' - A' B',
\end{align}
is to construct Bell's operators $(A,A',B,B')$ fulfilling:
\begin{align}\label{BellProp}
	A^2 = B^2 = 1; \quad A^\dagger = A, \, B^\dagger = B; \quad \left[A,B\right] = 0,
\end{align}
with similar equations holding for the primed operators. As we are working in Hilbert spaces which have infinite dimensions, this task can be accomplished by means of the pseudospin operators~\cite{psi1,psi2,psi3} defined as
\begin{equation} 
s_x = \sum_{n=0}^\infty s^{(n)}_x \;, \qquad s_y = \sum_{n=0}^\infty s^{(n)}_y \;, \qquad s_z = \sum_{n=0}^\infty s^{(n)}_z \;, \label{spin1} 
\end{equation}
where 
\begin{eqnarray} 
s^{(n)}_x & = & \vert 2n+1 \rangle \langle 2n \vert + \vert 2n \rangle \langle 2n+1 \vert, \nonumber \\ 
s^{(n)}_y & = &i\left( \vert 2n \rangle \langle 2n+1 \vert   - \vert 2n+1 \rangle \langle 2n \vert \right), \nonumber \\ 
s^{(n)}_z & = & \vert 2n+1 \rangle \langle 2n+1 \vert - \vert 2n \rangle \langle 2n \vert. \label{spin2}
\end{eqnarray} 
An easy calculation shows that 
\begin{align} 
\left[s^{(n)}_x,s^{(n)}_y \right] &= 2 i s^{(n)}_z \;, \nonumber \\
\left[s^{(n)}_y,s^{(n)}_z \right] &= 2 i s^{(n)}_x \;, \nonumber \\
\left[s^{(n)}_z,s^{(n)}_x \right] &= 2 i s^{(n)}_y \;.  \label{spin4}
\end{align} 
As a consequence, it follows that these operators obey the same algebraic relations of the spin $1/2$ Pauli matrices
\begin{align} 
	\left[ s_x,s_y \right] = 2 i s_z, \quad 	\left[s_y,s_z \right] = 2 i s_x, \quad 	\left[s_z,s_x \right] = 2 i s_y, \label{spin5}
\end{align} 
from which the name {\it pseudospin} follows. 

In particular, from expressions~\eqref{spin2} one observes that the introduction of the pseudospin operators can be related to a pairing mechanism in Hilbert space, a pair being given by two states, namely $(\vert2n \rangle, \vert2n+1\rangle)$, with $n=0,1,2,..$. Each pair of states gives raise to a set of operators, $(s^{(n)}_x,s^{(n)}_y,s^{(n)}_z)$, which obey the same spin $1/2$ algebra of Pauli matrices.  

The observation of the pairing mechanism  goes back to \cite{Gisin}. More recently, its applications to the study of the Bell-CHSH inequality has been discussed in \cite{Peruzzo:2023nrr,Sorella:2023hku,Sorella:2023iwz}, where it has been shown that each single pair might be employed for a test of the Bell-CHSH inequality.  This is the setup which we shall adopt in the following. 

More precisely, we shall analyze the Bell-CHSH inequality by considering two cases, namely: 

\begin{itemize} 
\item The Bell operators act non-trivially on a single pair, identified, for example, with the states $(\vert 0 \rangle, \vert 1 \rangle)$. Let $\vert x, y \rangle$ stand for a generic basis element of the Hilbert space ${\cal H} \otimes {\cal H}$ to which the entangled coherent states, Eqs.~\eqref{Symc},\eqref{AsymC},\eqref{Catc}, belong. Then, for the operators $(A,B)$ we shall set 
\begin{align}\label{BellOptwomodes}
	A \vert 0, y \rangle &= e^{i a}  \vert 1, y \rangle; \,\, 	A \vert 1, y \rangle = e^{-i a} \vert 0, y \rangle; \, \forall y, \nonumber \\
	B \vert x, 0 \rangle &= e^{i b}  \vert x, 1\rangle; \, 	B \vert x, 1 \rangle = e^{-i b} \vert x, 0 \rangle; \, \forall x, 
\end{align}
and acting as the identity on all the other states, {\it i.e.},
\begin{align}\label{BellOptwomodes2}
	A \vert x, y \rangle &= \vert x, y \rangle, \quad \forall x \geq 2, \nonumber \\
	B \vert x, y \rangle &= \vert x, y \rangle, \quad \forall y \geq 2.
\end{align}
The quantities $(a,b)$ are arbitrary real parameters. One sees that the operator $A$ acts only on the first entry of $\vert x, y \rangle$, while the operator $B$ only on the second one. In terms of the pseudospin operators, it turns out that the operator $A$ can be written as 
\begin{equation} 
A = \left( {\vec u} \cdot {\vec s}^{(0)} + {\cal R} \right) \otimes I,
\end{equation}
where $\vec{u}$ denotes the unit vector 
\begin{equation}
{\vec u} =\left( \cos(a), -\sin(a),0 \right), \qquad {\vec u} \cdot {\vec u} = 1, \label{vecu}
\end{equation} 
and ${\cal R}$ is the identity operator for $x\ge 2$:
\begin{equation} 
{\cal R} = \sum_{n=2}^\infty \vert n \rangle \langle n \vert.    \label{Rop}
\end{equation}
Analogous expressions can be  written down for $B$, $A'$ and $B'$. For the primed operators, the parameters $a$ and $b$ are simply replaced by $a'$ and $b'$. 
\item In the same vein, we can define a second Bell setup:
\begin{align}\label{BellOpallmodes}
	A \vert 2n, y \rangle &= e^{i a} \vert 2n + 1, y \rangle; \nonumber \\
	A \vert 2n + 1, y \rangle &= e^{-i a} \vert 2n, y \rangle; \, \forall y, \nonumber \\
	B \vert x, 2n \rangle &= e^{i b} \vert x, 2n + 1\rangle; \nonumber \\
	B \vert x, 2n + 1 \rangle &= e^{-i b} \vert x, 2n \rangle; \, \forall x.
\end{align}
with similar expressions for the primed operators $(A', B')$. One sees that, in the second setup, all states of the Hilbert space have been grouped  in pairs. In terms of pesusdospin operators, we have 
\begin{equation} 
A = {\vec u} \cdot {\vec s}   \otimes I,  \label{A2st}
\end{equation}
where $\vec{u}$ is the unit vector of expression \eqref{vecu}. 
\end{itemize}
It is immediate to check that in both setups the required properties~\eqref{BellProp} for the Bell-type operators are satisfied. 
We remark that in order to investigate the Bell-CHSH inequality violation, one needs only to compute the expected value of the product $AB$ on the state $\vert \psi \rangle$ of interest, since all the other combinations,  that is, $A'B$, $A B'$, and  $A' B'$, can be achieved by putting primes on the respective parameters. Therefore, in the following we will restrict ourselves to explicitly state only the result for $\langle \psi \vert A B \vert \psi \rangle$, being the complete expression for $\langle \psi \vert \mathcal{C}_{CHSH} \vert \psi \rangle$ left understood to avoid cluttering the text.

We are ready now to investigate the Bell-CHSH inequality violation for the three entangled coherent states~\eqref{Symc},\eqref{AsymC} and \eqref{Catc}, starting with the symmetric state.

\section{Symmetric coherent states}\label{SecSymmCohe}

Let us consider here the  symmetric entangled coherent states, a generalization of the states considered in~\cite{Gerry09}, obtained by the entanglement of two different coherent states with an arbitrary relative phase $\phi$:  
\begin{align}\label{SymmCoherent}
	\vert \psi \rangle_S = N_S \left[ \vert \alpha \rangle_a \vert \beta \rangle_b + e^{i \phi} \vert \beta \rangle_a \vert \alpha \rangle_b \right],
\end{align}
where the normalization factor  is given by
\begin{align}\label{key}
	N_S = \frac{1}{\sqrt{2}} \left(1+ \cos\phi \exp\left[-\left(\alpha -\beta \right)^2\right]\right)^{-1/2}.
\end{align}
In the sequel, we analyze the Bell-CHSH violation for the symmetric states~\eqref{SymmCoherent} considering the two different Bell setups defined  in Eqs.~\eqref{BellOptwomodes},\eqref{BellOptwomodes2},\eqref{BellOpallmodes}.

\begin{figure}[t!]
	\begin{minipage}[b]{1.0\linewidth}
		\includegraphics[width=\textwidth]{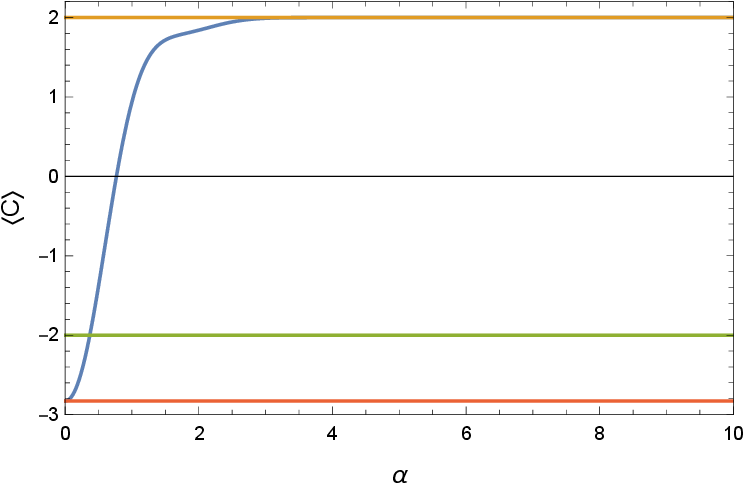}
	\end{minipage} \hfill
	\caption{$\langle \mathcal{C} \rangle$ correlator in blue, as a function of $\alpha$ in the symmetric case (first setup), with $\beta = \alpha + 0.001$. Here we considered $\phi = \pi, \, a = 0,\, a' = \pi/2,\, b = +\pi/4,\, b' = -\pi/4$. The red line represents the Tsirelson's bound.}
	\label{BellSymmTwo2D}
\end{figure}
\begin{figure}[t!]
	\begin{minipage}[b]{1.0\linewidth}
		\includegraphics[width=\textwidth]{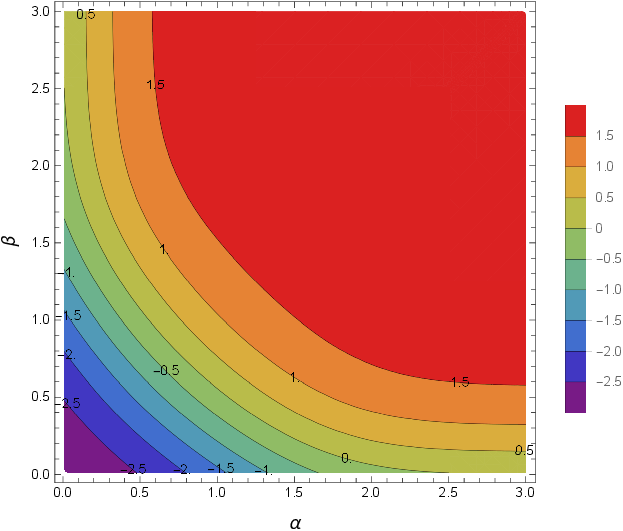}
	\end{minipage} \hfill
	\caption{Contour plot exhibiting $ \langle \mathcal{C} \rangle$ as a function of $\alpha$ and $\beta$ in the symmetric case (first setup). Here we have phase $\phi = \pi$, and parameters $a = 0,\, a' = \pi/2,\, b = +\pi/4,\, b' = -\pi/4$.  There is violation whenever we have $\vert \langle \mathcal{C} \rangle \vert > 2$.}
	\label{BellSymmTwoContour}
\end{figure}

\subsection{First setup}\label{Symm1}

Adopting the first setup, defined by Eqs.~\eqref{BellOptwomodes},\eqref{BellOptwomodes2}, we can compute the expectation value $\langle \psi \vert AB \vert \psi \rangle$ in the symmetric state defined by Eq.~\eqref{SymmCoherent}. Thus, we find:

\begin{align}\label{key}
	\langle A B \rangle_{S_1} &= \Omega_{S} \Big\{ \left(\cos a + \cos b\right) \big[ \beta \left(e^{\alpha^2} -1 -\alpha^2 \right) \nonumber \\
	&+ \alpha \left(e^{\beta^2} -1 - \beta^2\right)\big] + \left(e^{\alpha \beta} - 1 -\alpha \beta\right) \nonumber \\
	&\times\big[ \alpha \left(\cos(a + \phi) + \cos(b - \phi)\right) \nonumber \\
	&+ \beta \left(\cos(a - \phi) + \cos(b + \phi)\right)\big] \nonumber \\
	&+ \big[4 \alpha \beta \cos a \cos b + 2 \alpha \beta \cos\phi \cos(a+b) \nonumber \\
	&+ \beta^2 \cos(a - b -\phi) + \alpha^2 \cos(a - b + \phi)       \big] \nonumber \\
	&+\big[\left( e^{\alpha^2} - 1 -\alpha^2 \right) \left( e^{\beta^2} -1 -\beta^2 \right)  \nonumber \\
	&+ \cos\phi \left(e^{\alpha \beta} - 1 - \alpha \beta\right)^2 \big] \Big\},
\end{align}
where the overall factor is given by
\begin{align}\label{OmegaS}
	\Omega_{S} = \frac{\exp\left[-\left(\alpha^2 + \beta^2\right)\right]}{1 + \cos\phi \, \exp\left[-\left(\alpha - \beta\right)^2\right]}.
\end{align}

Considering a relative phase $\phi = \pi$, we will adopt the following choice of parameters to find a Bell-CHSH inequality violation: $a = 0,\, a' = \pi/2,\, b = +\pi/4,\, b' = -\pi/4$. In this case, the expression for $\langle \mathcal{C}_{CHSH} \rangle$ simplifies considerably, giving us:
\begin{align}\label{key}
	\langle \mathcal{C}_{CHSH} \rangle &= \frac{e^{-(\alpha^2 + \beta^2)}}{1 - e^{-(\alpha - \beta)^2}} \Big\{ 2 e^{\alpha^2 + \beta^2} \left(1 - e^{-(\alpha - \beta)^2}\right) \nonumber \\
	&- 2 (\sqrt{2} -1) (\alpha-\beta)^2 - e^{\beta^2}\left[2 - (2+\sqrt{2})\alpha + 2 \alpha^2\right] \nonumber \\
	&- e^{\alpha^2}\left[2 - (2+\sqrt{2})\beta + 2 \beta^2\right] \nonumber \\
	&+e^{\alpha \beta}\left[4 + 4 \alpha \beta - (2 + \sqrt{2}) (\alpha + \beta)\right]\Big\}.
\end{align}
From the above expression, taking $\alpha = 0.1$ and $\beta=0.2$ we already obtain $\vert\langle \mathcal{C}_{CHSH} \rangle\vert = 2.6939$. Interestingly, if we keep one of them small, we can take the other large and still find a violation. For instance, keeping $\beta = 0.01$, we can take $\alpha = 0.70$ and still find $\vert\langle \mathcal{C} \rangle \vert= 2.1699$. Considering smaller $\alpha$ and $\beta$, for instance, $(\alpha, \beta) = (0.0001, 0.0002)$, we can saturate the Tsirelson's bound finding $\vert\langle \mathcal{C} \rangle \vert= 2.8284 \approx 2 \sqrt{2}$. It is important to remark that with the parameters adopted here, the result is symmetric under the exchange $\alpha \leftrightarrow \beta$. 
Finally, considering fixed values for $\alpha$ and $\beta$, we can find a small range for the relative phase around $\phi=\pi$ that still exhibits a violation. For instance, taking $(\alpha, \beta) = (0.5, 0.1)$, we still observe a violation for phases $\phi$ in the range $(\pi - 0.2, \pi+2)$. 
Without a relative phase ($\phi=0$), we were not able to find any violation.

In order to simplify the visualization of these features, we show $\langle \mathcal{C} \rangle$ as a function of $\alpha$, considering $\beta = \alpha + 0.001$ in Fig.~\ref{BellSymmTwo2D}. Notice that there is violation for small values of $\alpha$, a saturation of the Tsirelson's bound for very small $\alpha$, and the result asymptotes to $2$ for large values of $\alpha$, a feature that can already be seen  around $\alpha \approx 3$. One can see explicitly the region of parameters $(\alpha, \beta)$ leading to Bell-CHSH inequality violation in Fig.~\ref{BellSymmTwoContour}.

\subsection{Second setup}\label{Symm2}

Now, continuing our analysis, we consider the more general setup given by Eq.~\eqref{BellOpallmodes}. In this case, we find for the expectation value of $AB$ in the symmetric state:
\begin{align}\label{key}
	\langle A B \rangle_{S_2} &= \Omega_{S} \sum_{n,m=0}^{\infty} \frac{1}{\sqrt{(2n)! (2n+1)! (2m)! (2m+1)!}}  \nonumber \\
	&\times\Big\{ 4 \cos a  \cos b \, \alpha^{4n+1} \beta^{4m+1}     \nonumber \\
	&+  (\alpha \beta)^{2n + 2m} \big[ 2 \alpha \beta \cos\phi \, \cos(a + b) \nonumber \\
	&+ \alpha^2 \cos(a - b + \phi) + \beta^2 \cos(a - b -\phi)\big] \Big\},
\end{align}
where the overall factor here is the same as before, given by Eq.~\eqref{OmegaS}.
This time, unfortunately, we were not able to find a closed analytical expression for the above sum. Even tough, we can do the analysis by considering the appropriate number of terms in the series in order to stabilize the result for a given choice of parameters.

Once again, we consider $\phi=\pi$ and the following choice of parameters: $a = 0,\, a' = \pi/2,\, b = +\pi/4,\, b' = -\pi/4$. First of all, we take the first contribution in the series, that is, the $n=m=0$ term. It can be written as: 
\begin{align}\label{key}
 \langle A B \rangle\vert_{n=m=0} = \frac{2 \sqrt{2} (\alpha -\beta )^2}{e^{2 \alpha  \beta }-e^{\alpha ^2+\beta ^2}}.
\end{align}
For small values of $\alpha$ and $\beta$, the first order term is already $-2 \sqrt{2}$ plus small corrections, encouraging us to proceed and consider the complete expression. Upon considering $(\alpha, \beta) = (0.1, 0.2)$, we already find $\vert\langle \mathcal{C}_{CHSH} \rangle \vert=  2.7018$. As before, we can saturate the Tsirelson's bound if we take smaller values of $\alpha$ and $\beta$. Once again, keeping one of them small, we have freedom to enlarge the other and still observe a violation. For instance, taking $(\alpha, \beta) = (0.70, 0.001)$, we obtain $\vert\langle \mathcal{C}_{CHSH} \rangle \vert=  2.1732$. Upon choosing $(\alpha, \beta) = (0.5, 0.1)$ as before, we still observe a violation for phases $\phi$ in the range $(\pi - 0.2, \pi+2)$, and we were not able to find any violation for $\phi=0$.

\begin{figure}[t!]
	\begin{minipage}[b]{1.0\linewidth}
		\includegraphics[width=\textwidth]{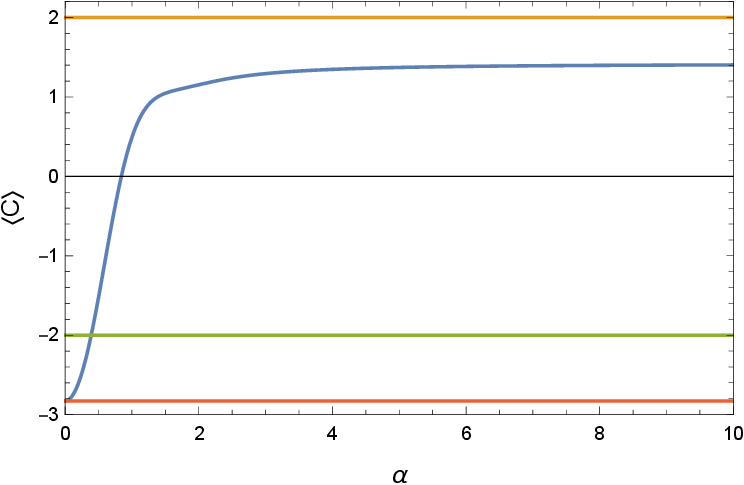}
	\end{minipage} \hfill
	\caption{$\langle \mathcal{C} \rangle$ correlator in blue, as a function of $\alpha$ in the symmetric case (second setup), with $\beta = \alpha + 0.001$. Here we considered $\phi = \pi, \, a = 0,\, a' = \pi/2,\, b = +\pi/4,\, b' = -\pi/4$. The red line represents the Tsirelson's bound.}
	\label{BellSymmAll2D}
\end{figure}

\begin{figure}[t!]
	\begin{minipage}[b]{1.0\linewidth}
		\includegraphics[width=\textwidth]{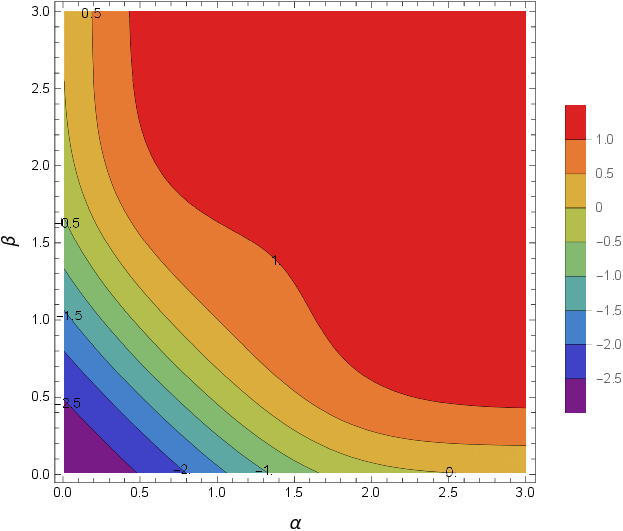}
	\end{minipage} \hfill
	\caption{Contour plot exhibiting $ \langle \mathcal{C} \rangle$ as a function of $\alpha$ and $\beta$ in the symmetric case (second setup). Here we have  phase $\phi = \pi$, and parameters $a = 0,\, a' = \pi/2,\, b = +\pi/4,\, b' = -\pi/4$.  There is violation whenever we have $\vert \langle \mathcal{C} \rangle \vert > 2$.}
	\label{BellSymmAllContour}
\end{figure}

As before, we show $\langle \mathcal{C} \rangle$ as a function of $\alpha$, considering $\beta = \alpha + 0.001$ in Fig.~\ref{BellSymmAll2D}. The region of parameters $(\alpha, \beta)$ leading to Bell-CHSH inequality violation can be seen in Fig.~\ref{BellSymmAllContour}. We remark that the qualitative behavior of these graphs are the same as in the last subsection, with only small quantitative differences between them.


\section{Asymmetric coherent states}\label{SecAsymmCohe}

Now, we consider the asymmetric entangled coherent states, similar to the case analyzed in~\cite{Park15}, but with a different Bell setup, and here considering a general relative phase $\phi$. Thus, we define these states as 
\begin{align}\label{AsymmCoherent}
\vert \psi \rangle_A = N_A \left[ \vert \alpha \rangle_a \vert \beta \rangle_b + e^{i \phi} \vert -\alpha \rangle_a \vert -\beta \rangle_b \right],
\end{align}
where the normalization factor here is given by 
\begin{align}\label{key}
	N_A = \frac{1}{\sqrt{2}} \left(1+ \cos\phi \exp\left[-2\left(\alpha^2+\beta^2\right)\right]\right)^{-1/2}.
\end{align}
In the following, we analyze the Bell-CHSH violation for the asymmetric states~\eqref{AsymmCoherent} considering the two  Bell setups defined before, as was done for the symmetric states.

\vspace{-0.2cm}
\subsection{First setup}\label{Asymm1}

To begin with, we consider the first setup defined by Eqs.~\eqref{BellOptwomodes},\eqref{BellOptwomodes2}.
Computing the expectation value  $\langle \psi \vert A B \vert \psi \rangle$ on the asymmetric state defined in Eq.~\eqref{AsymmCoherent}:
\begin{align}\label{key}
\langle A B \rangle_{A_1} &= \Omega_{A} \Big\{ 4 \alpha \beta \left(  \cos a \cos b - \cos\phi \sin a \sin b \right) \nonumber \\
&- 2 \sin\phi \big[ \alpha \sin a \left(-1 + \beta^2 + e^{-\beta^2}\right) \nonumber \\
&+ \beta \sin b \left(-1 + \alpha^2 + e^{-\alpha^2}\right)\big] \nonumber \\
&+\big[ \left(e^{\alpha^2} - 1 - \alpha^2\right) \left(e^{\beta^2} - 1 - \beta^2\right) \nonumber \\
&+ \cos\phi  \left(e^{-\alpha^2} - 1 + \alpha^2\right) \left(e^{-\beta^2} - 1 + \beta^2\right)\big]\Big\},
\end{align}
where the overall factor here is given by 
\begin{align}\label{OmegaA}
	\Omega_{A} = \frac{\exp\left[-\left(\alpha^2 + \beta^2\right)\right]}{1+ \cos\phi \exp\left[-2\left(\alpha^2 + \beta^2\right)\right]}.
\end{align}

Here we also consider the relative phase $\phi=\pi$ and the same set of parameters used in the last section, that is: $a = 0,\, a' = \pi/2,\, b = +\pi/4,\, b' = -\pi/4$. With this choice, the expression for $\langle \mathcal{C} \rangle$ significantly simplifies:
\begin{align}\label{key}
&\langle \mathcal{C}_{CHSH} \rangle = 2 - \frac{2}{\sinh(\alpha^2 + \beta^2)} \Big[-2 \sqrt{2} \alpha  \beta  - \alpha^2 -\beta^2   \nonumber \\ 
&+\sinh \alpha ^2 +\sinh \beta ^2 + \alpha^2 \cosh \beta ^2 +\beta^2 \cosh \alpha ^2\Big].
\end{align}
From the above expression one can see that there is a considerable range of parameters $\alpha$ and $\beta$ providing a violation. In fact, we observe that the Tsirelson's bound is saturated already with $\alpha = \beta = 0.06$.  Moreover, upon keeping $\alpha = \beta$, we can find violations from $\alpha = \beta = 0.1$ (with $\langle \mathcal{C}_{CHSH} \rangle = 2.8282$) to $\alpha = \beta = 0.8$ (with  $\langle \mathcal{C} \rangle = 2.0218$). But if we do not take them close to each other, we do not find any violation. For instance, considering $(\alpha, \beta)=(0.1, 0.3)$ we find $\langle \mathcal{C}_{CHSH} \rangle = 1.6942$ while for $(\alpha, \beta)=(0.3, 0.3)$ we find $\langle \mathcal{C}_{CHSH} \rangle = 2.8132$. Therefore, in this case it is important to keep $\alpha$ and $\beta$ close to find violations. As before, fixing values for $\alpha$ and $\beta$, we can search for violation with relative phases around $\phi = \pi$. In the asymmetric case, we have more space for violation in the relative phase than in the previous case. In fact, adopting $\alpha = \beta = 0.5$, we find violation for $\phi \in (\pi-0.78, \pi + 0.78)$, a considerably larger range in comparison with the symmetric case studied before.

\begin{figure}[t!]
	\begin{minipage}[b]{1.0\linewidth}
		\includegraphics[width=\textwidth]{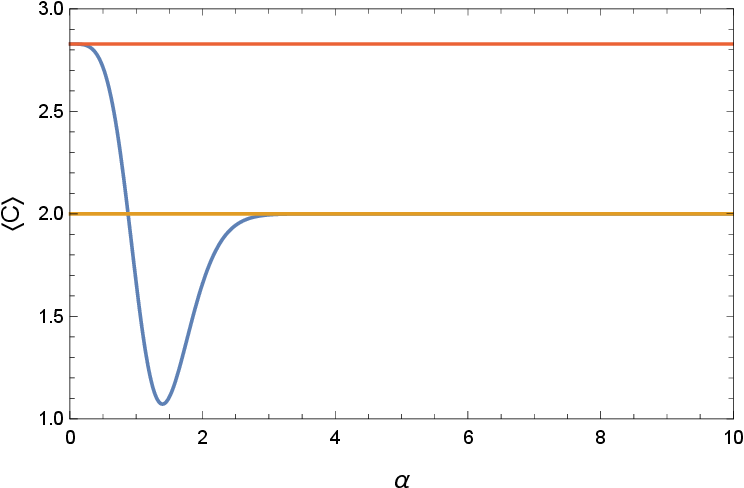}
	\end{minipage} \hfill
	\caption{$\langle \mathcal{C} \rangle$ correlator in blue, as a function of $\alpha$ in the asymmetric case (first setup), with $\beta = \alpha$. Here we considered $\phi = \pi, \, a = 0,\, a' = \pi/2,\, b = +\pi/4,\, b' = -\pi/4$. The red line represents the Tsirelson's bound.}
	\label{BellAsymmTwo2D}
\end{figure}

\begin{figure}[t!]
	\begin{minipage}[b]{1.0\linewidth}
		\includegraphics[width=\textwidth]{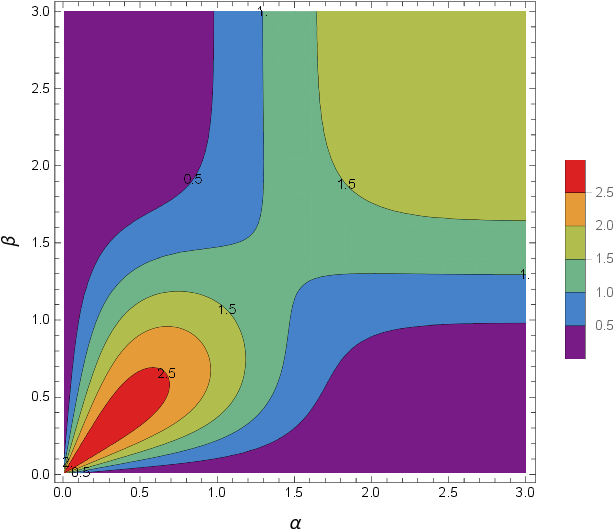}
	\end{minipage} \hfill
	\caption{Contour plot exhibiting $ \langle \mathcal{C} \rangle$ as a function of $\alpha$ and $\beta$ in the asymmetric case (first setup). Here we have phase $\phi = \pi$, and parameters $a = 0,\, a' = \pi/2,\, b = +\pi/4,\, b' = -\pi/4$.  There is violation whenever we have $\vert \langle \mathcal{C} \rangle \vert > 2$.}
	\label{BellAsymmTwoContour}
\end{figure}

In order to visualize better the results, we show $\langle \mathcal{C} \rangle$ as a function of $\alpha$ in Fig.~\ref{BellAsymmTwo2D}, considering $\alpha = \beta$. Notice that the region of parameters $\alpha$ leading to a violation is considerably larger than the one obtained in the symmetric case studied before. Moreover, we saturate the Tsirelson's bound for small $\alpha$, and the result asymptotes to $2$ for large values of $\alpha$, something that can already be seen around $\alpha \approx 3$, as before. The region of parameters $(\alpha, \beta)$ leading to a violation is shown in Fig.~\ref{BellAsymmTwoContour}. We also remark that there is a more pronounced violation in the asymmetric case, comparing with the symmetric one.


\subsection{Second setup}\label{Asymm2}

Giving sequence to our analysis, we now consider the more general setup defined by Eq.~\eqref{BellOpallmodes}. The expectation value $\langle \psi \vert A B \vert \psi \rangle$ in the asymmetric state defined in Eq.~\eqref{AsymmCoherent} is given  by:
\begin{align}\label{key}
	\langle A B &\rangle_{A} = 4 \Omega_{A_2} \, \left( \cos a \cos b - \cos\phi \sin a \sin b\right) \nonumber \\
	&\times\sum_{n,m=0}^{\infty}\left[\frac{\alpha^{4n+1} \beta^{4m+1}}{\sqrt{(2n)! (2n+1)! (2m)! (2m+1)!}}\right],
\end{align}
where the overall factor here is the same as before, given by Eq.~\eqref{OmegaA}.
Although the expression obtained is pretty simple, this infinite sum cannot be written in a closed analytic form. The simple angular structure of this expression will make this example very interesting.

Also here, we choose to work with the same parameters $a = 0,\, a' = \pi/2,\, b = +\pi/4,\, b' = -\pi/4$. Firstly, we consider the first contribution of this sum, taking the $n=m=0$ term, that can be written as:
\begin{align}\label{key}
	\langle A B \rangle \vert_{n=m=0} = \frac{4 \sqrt{2} \, \alpha  \beta}{\sinh\left(\alpha ^2+\beta ^2\right)}.  
\end{align}
For small values of $\alpha $ and $\beta$ we have a result close to $2\sqrt{2}$, encouraging us to proceed with the analysis using the complete expression. Taking $\alpha = \beta = 0.1$, we already find $\langle \mathcal{C}_{CHSH}\rangle = 2.8284 \simeq 2 \sqrt{2}$. Once again, for small values of $\alpha$ and $\beta$, it is important to keep these parameters close to each other in order to find a violation. Considering $\alpha=\beta$, we have found a Bell-CHSH inequality violation for all the considered parameters $\alpha$. For instance, taking $\alpha = \beta = 0.01$ we find $\langle \mathcal{C}_{CHSH} \rangle = 2.8284$; for $\alpha = \beta = 1$, we have $\langle \mathcal{C} \rangle = 2.6678$; considering $\alpha = \beta = 5$, we have $\langle \mathcal{C} \rangle = 2.7997$. We remark that here we also have a considerable freedom in the phase. For example, adopting $\alpha = \beta = 0.5$, we find violation for any $\phi \in (\pi - 0.81, \pi  +0.81)$. 

\begin{figure}[t!]
	\begin{minipage}[b]{1.0\linewidth}
		\includegraphics[width=\textwidth]{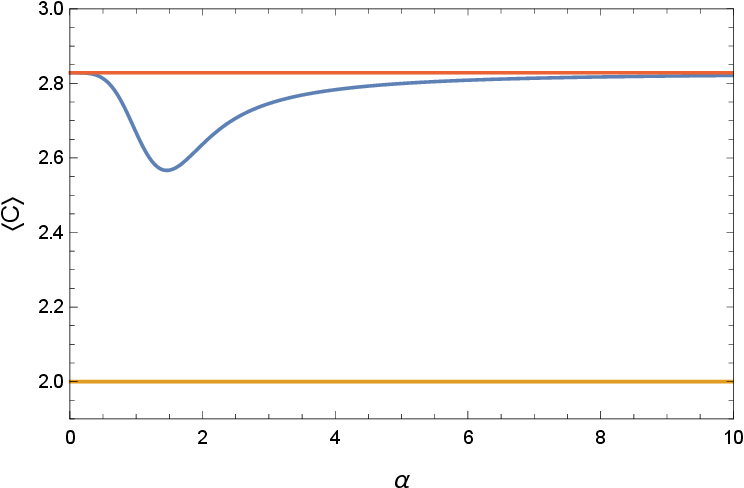}
	\end{minipage} \hfill
	\caption{$\langle \mathcal{C} \rangle$ correlator in blue, as a function of $\alpha$ in the asymmetric case (second setup), with $\beta = \alpha$. Here we considered $\phi = \pi, \, a = 0,\, a' = \pi/2,\, b = +\pi/4,\, b' = -\pi/4$. The red line represents the Tsirelson's bound.}
	\label{BellAsymmAll2D}
\end{figure}

\begin{figure}[t!]
	\begin{minipage}[b]{1.0\linewidth}
		\includegraphics[width=\textwidth]{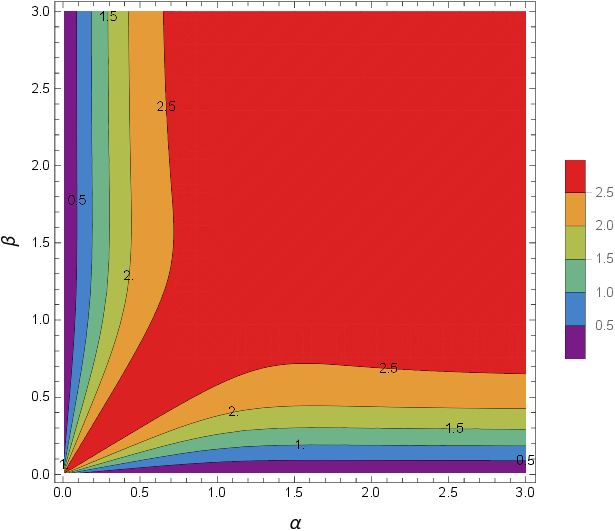}
	\end{minipage} \hfill
	\caption{Contour plot exhibiting $ \langle \mathcal{C} \rangle$ as a function of $\alpha$ and $\beta$ in the asymmetric case (second setup). Here we have phase $\phi = \pi$, and parameters $a = 0,\, a' = \pi/2,\, b = +\pi/4,\, b' = -\pi/4$.  There is violation whenever we have $\vert \langle \mathcal{C} \rangle \vert > 2$.}
	\label{BellAsymmAllContour}
\end{figure}

Once more, we show $\langle \mathcal{C} \rangle$ as a function of $\alpha$ in Fig.~\ref{BellAsymmAll2D}, considering $\alpha=\beta$. In this case, we are finding a large violation for all the considered values of $\alpha$. The region of parameters $(\alpha, \beta)$ leading to a violation is also large, as one can immediately see in Fig.~\ref{BellAsymmAllContour}.


\begin{figure}[t!]
	\begin{minipage}[b]{1.0\linewidth}
		\includegraphics[width=\textwidth]{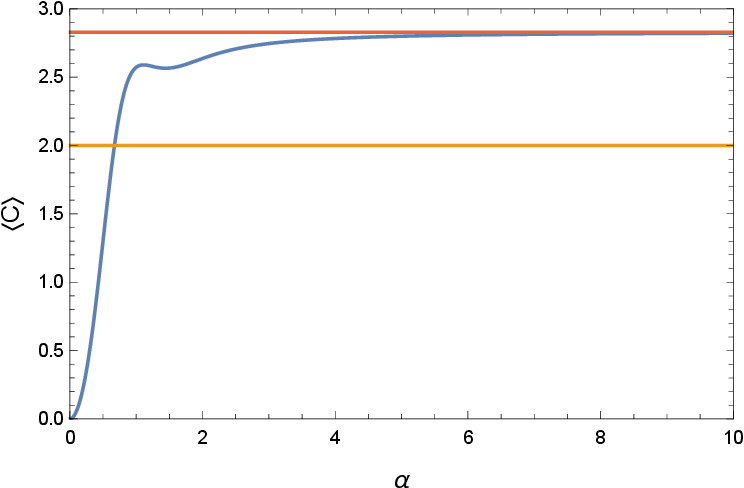}
	\end{minipage} \hfill
	\caption{$\langle \mathcal{C} \rangle$ correlator in blue, as a function of $\alpha$ in the asymmetric case (second setup), with $\beta = \alpha$. Here we considered $\phi = 0, \, a = 0,\, a' = \pi/2,\, b = +\pi/4,\, b' = -\pi/4$. The red line represents the Tsirelson's bound.}
	\label{BellAsymmWithout2D}
\end{figure}

\begin{figure}[t!]
	\begin{minipage}[b]{1.0\linewidth}
		\includegraphics[width=\textwidth]{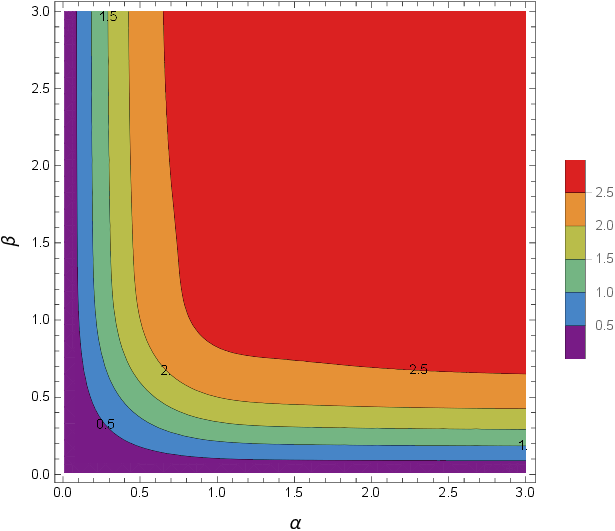}
	\end{minipage} \hfill
	\caption{Contour plot exhibiting $ \langle \mathcal{C} \rangle $ as a function of $\alpha$ and $\beta$ in the asymmetric case (second setup). Here we have  phase $\phi = 0$, and parameters $a = 0,\, a' = \pi/2,\, b = -\pi/4,\, b' = +\pi/4$.  There is violation whenever we have $\vert \langle \mathcal{C} \rangle \vert > 2$.}
	\label{BellAsymmWithoutContour}
\end{figure}

Finally, it is noteworthy that in this case, we managed to find a Bell-CHSH inequality violation even in the case without a relative phase, that is, with $\phi=0$. In fact, taking $\phi=0$ we can consider now a slightly different set of parameters: $a = 0,\, a' = \pi/2,\, b = -\pi/4,\, b' = +\pi/4$. Thus, for $\alpha=\beta=0.7$ we find already $\langle \mathcal{C} \rangle = 2.0895$. There is violation in this case also for larger values of $\alpha$, as one can see in Fig.~\ref{BellAsymmWithout2D}. For different values of $\alpha$ and $\beta$, see Fig.~\ref{BellAsymmWithoutContour}. We remark that a Bell-CHSH inequality violation for the states defined by Eq.~\eqref{AsymmCoherent} for $\phi=\pi$ and $\phi=0$ was also obtained in~\cite{Park15}, although with different Bell setups.

\section{Schr\"odinger's Cat states}\label{SecCatCohe}

Let us proceed now with  the so-called entangled Schr\"odinger cat states.
One can define both the even $(+)$ and odd $(-)$ cases simultaneously by the equation:
\begin{align}\label{key}
	\vert \alpha \rangle_\pm = N_\pm \left[ \vert \alpha \rangle \pm \vert -\alpha \rangle \rangle \right],
\end{align}
with the normalizations respectively given by
\begin{align}\label{key}
	N_+ = \frac{e^{\alpha^2 / 2}}{2 \sqrt{\cosh(\alpha^2)}}, \quad 	N_- = \frac{e^{\alpha^2 / 2}}{2 \sqrt{\sinh(\alpha^2)}}.
\end{align}
Therefore, we can consider the two entangled states, considering the even and odd cases separately:
\begin{align}\label{CatStates}
	\vert \psi \rangle_\pm = C_\pm \left[\vert \alpha \rangle_\pm \vert \beta \rangle_\pm + e^{i \phi}  \vert \beta \rangle_\pm \vert \alpha \rangle_\pm\right],
\end{align}
with normalizations in this case given respectively by
\begin{align}\label{key}
	C_+ &= \frac{1}{\sqrt{2}} \left[1+ \cos\phi \frac{\cosh^2(\alpha \beta)}{\cosh(\alpha^2) \cosh(\beta^2)}\right]^{-1/2}, \\
	C_- &= \frac{1}{\sqrt{2}} \left[1+ \cos\phi \frac{\sinh^2(\alpha \beta)}{\sinh(\alpha^2) \sinh(\beta^2)}\right]^{-1/2}.
\end{align}

In this case, we need to adapt a bit our Bell setup, since for the even cat state there are only even modes $\vert 2n \rangle$ and for the odd cat states there are only odd modes $\vert 2n +1 \rangle$. Thus, in this section, we slightly modify our first setup for the even cat states, using
\begin{align}\label{CatTwoModesEven}
	A \vert 0, y \rangle &= e^{i a}  \vert 2, y \rangle; \,\, 	A \vert 2, y \rangle = e^{-i a} \vert 0, y \rangle; \, \forall y, \nonumber \\
	B \vert x, 0 \rangle &= e^{i b}  \vert x, 2\rangle; \, 	B \vert x, 2 \rangle = e^{-i b} \vert x, 0 \rangle; \, \forall x, 
\end{align}
and also for the odd cat states, by defining
\begin{align}\label{CatTwoModesOdd}
	A \vert 1, y \rangle &= e^{i a}  \vert 3, y \rangle; \,\, 	A \vert 3, y \rangle = e^{-i a} \vert 1, y \rangle; \, \forall y, \nonumber \\
	B \vert x, 1 \rangle &= e^{i b}  \vert x, 3\rangle; \, 	B \vert x, 3 \rangle = e^{-i b} \vert x, 1 \rangle; \, \forall x, 
\end{align}
and acting as the identity in all other states in both setups. Furthermore, since the cat states $\vert \alpha \rangle_+$ and $\vert \alpha \rangle_-$ have only even and odd modes, respectively, the second Bell setup defined earlier would give vanishing expectation value in this case. Therefore, in the following, we will not consider it, limiting ourselves to the Bell-CHSH inequality violation analysis for the setup described above in this section, for both even and odd entangled cat states defined by Eq.~\eqref{CatStates}.


\subsection{Even Cat States}

First, we consider the even cat state $\vert \psi \rangle_+$ as defined in Eq.~\eqref{CatStates}. For this state, one can compute the expectation value of the product $A B$ and find:
\begin{align}\label{key}
	&\langle A  B \rangle_+ = \Omega_+ \Bigg\{ 4 \alpha^2 \beta^2 \cos a \cos b + 2 \alpha^2 \beta^2 \cos\phi \cos(a+b) \nonumber \\
	&+\alpha^4 \cos(a-b+\phi) + \beta^4 \cos(a-b-\phi) \nonumber \\
	&+\sqrt{2} \left(\cos a + \cos b\right) \Big[\alpha^2 \left(\cosh(\beta^2) - 1 - \frac{\beta^4}{2}\right) \nonumber \\
	&+ \beta^2 \left(\cosh(\alpha^2) - 1 - \frac{\alpha^4}{2}\right)  \Big] \nonumber \\
	&+\sqrt{2} \Big[\alpha^2 \left(\cos(a+\phi) + \cos(b-\phi)\right) + \nonumber \\
	&+\beta^2 \left(\cos(a-\phi) + \cos(b+\phi)\right)\Big] \left(\cosh(\alpha \beta) - 1 -\frac{(\alpha \beta)^2}{2}\right)      \nonumber \\
	&+ 2 \left( \cosh(\alpha^2) - 1 -\frac{\alpha^4}{2}\right) \left(\cosh(\beta^2) - 1 -\frac{\beta^4}{2}\right) \nonumber \\
	&+ 2 \cos\phi \left(\cosh(\alpha \beta) - 1 -\frac{\alpha^2 \beta^2}{2}\right)^2 \Bigg\},
\end{align}
where the overall factor can be written as
\begin{align}\label{key}
	\Omega_+ = \frac{1}{2} \left[ \cosh(\alpha^2) \cosh(\beta^2) + \cos\phi \, \cosh^2(\alpha \beta)\right]^{-1}.
\end{align}

Once more, we adopt the phase $\phi=\pi$ and the usual set of parameters  $a = 0,\, a' = \pi/2,\, b = +\pi/4,\, b' = -\pi/4$. In this case, the expression for $\langle \mathcal{C}_{CHSH} \rangle$ significantly simplifies: 
\begin{align}\label{key}
	\langle \mathcal{C}_{CHSH} \rangle &= \frac{1}{\kappa_+} \Big\{ 1 + \left(\sqrt{2}-1\right) \left(\alpha ^2-\beta ^2\right)^2  \nonumber \\
	&+ \cosh(2 \alpha  \beta) -2 \cosh(\alpha^2) \cosh(\beta^2) \nonumber \\
	&- \cosh(\alpha \beta) \left[4 + 2\alpha^2 \beta^2 - (1+\sqrt{2}) (\alpha^2+ \beta^2)\right]    \nonumber \\
	&+\cosh(\alpha^2) \left[2 - (1+\sqrt{2}) \beta^2 + \beta^4\right] \nonumber \\
	&+ \cosh(\beta^2) \left[2 - (1+\sqrt{2}) \alpha^2 +  \alpha^4\right] \Big\},
\end{align}
where $\kappa_+ = \cosh^2(\alpha  \beta) - \cosh(\alpha^2) \cosh(\beta^2)$. Considering $(\alpha, \beta) = (0.1, 0.2)$ we find $\vert \langle \mathcal{C} \rangle \vert = 2.8278$ and the Tsirelson's bound is saturated already at $(\alpha, \beta) = (0.07, 0.08)$.  We remark that with the parameters adopted, the result is also symmetric under the exchange $\alpha \leftrightarrow \beta$ and that one finds a Bell-CHSH violation for almost the whole interval of  $\alpha$ and $\beta$ between $0$ and $1$.
Considering the parameters $(\alpha, \beta) = (0.1, 0.8)$, we still find violation for relative phases around $\phi \in (\pi - 0.3, \pi + 0.3)$. 

\begin{figure}[t!]
	\begin{minipage}[b]{1.0\linewidth}
		\includegraphics[width=\textwidth]{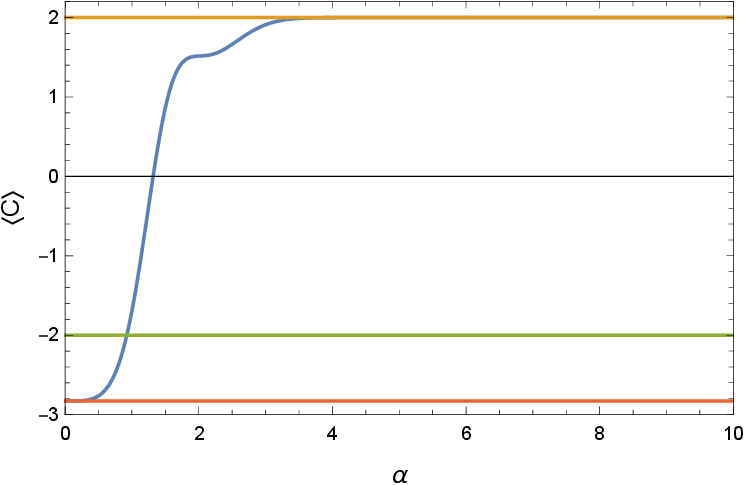}
	\end{minipage} \hfill
	\caption{$\langle \mathcal{C} \rangle$ correlator in blue, as a function of $\alpha$ in the even cat state, with $\beta = \alpha + 0.001$. Here we considered $\phi = \pi, \, a = 0,\, a' = \pi/2,\, b = +\pi/4,\, b' = -\pi/4$. The red line represents the Tsirelson's bound.}
	\label{BellCatPlus2D}
\end{figure}

\begin{figure}[t!]
	\begin{minipage}[b]{1.0\linewidth}
		\includegraphics[width=\textwidth]{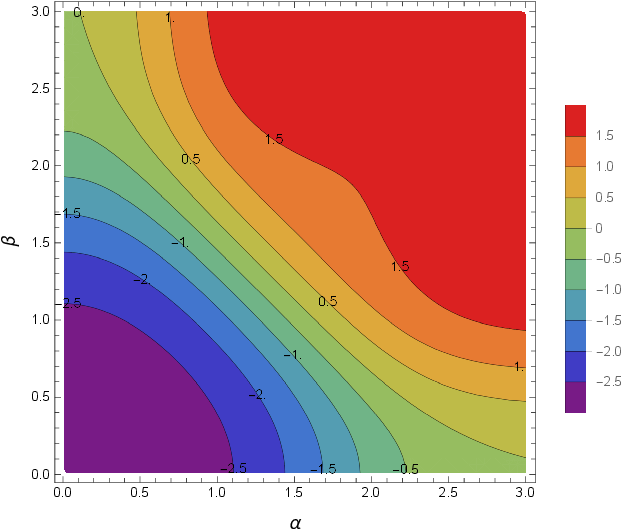}
	\end{minipage} \hfill
	\caption{Contour plot exhibiting $ \langle \mathcal{C} \rangle$ as a function of $\alpha$ and $\beta$ in the even cat state. Here we have phase $\phi = \pi$, and parameters $a = 0,\, a' = \pi/2,\, b = +\pi/4,\, b' = -\pi/4$. There is violation whenever we have $\vert \langle \mathcal{C} \rangle \vert > 2$.}
	\label{BellCatPlusContour}
\end{figure}

We exhibit $\langle \mathcal{C} \rangle$ as a function of $\alpha$ in Fig.~\ref{BellCatPlus2D}, taking $\beta = \alpha + 0.001$. Notice that there is violation for almost all values of $\alpha$ between $0$ and $1$, and that for larger values of $\alpha$ the result asymptotes to $2$, a behavior that can already be seen around $\alpha \approx 3$. For the range of parameters $(\alpha, \beta)$ leading to a violation, see Fig.~\ref{BellCatPlusContour}.


\subsection{Odd Cat States}

Continuing our analysis, we consider now the odd cat state $\vert \psi \rangle_-$ defined in Eq.~\eqref{CatStates}. For this state, the expectation value of the product $A B$ is given by:
\begin{align}\label{key}
	&\langle A  B \rangle_- = \Omega_- \Bigg\{ \frac{4}{3} \alpha^4 \beta^4 \cos a \cos b + \frac{2}{3} \cos \phi \, \alpha^4 \beta^4 \cos(a+b) \nonumber \\
	&+ \frac{1}{3} \alpha^2 \beta^2 \Big[ \alpha^4 \cos(a-b+\phi) + \beta^4 \cos(a-b-\phi)\Big] \nonumber \\
	&+ \frac{2}{\sqrt{6}} \left(\cos a + \cos b\right) \Big[\alpha^4 \left(\sinh(\beta^2) - \beta^2 - \frac{\beta^6}{6}\right) \nonumber \\
	&+ \beta^4 \left(\sinh(\alpha^2) - \alpha^2 - \frac{\alpha^6}{6}\right)\Big]   \nonumber \\
	&+ \frac{2}{\sqrt{6}}  \alpha \beta  \Big[\alpha^2 \left(\cos(a+\phi) + \cos(b-\phi)\right)  \nonumber \\
	&  + \beta^2 \left(\cos(a-\phi) + \cos(b+\phi)\right)\Big] \left(\sinh(\alpha \beta) - \alpha \beta - \frac{(\alpha \beta)^3}{6}\right) \nonumber \\
	&+2 \left(\sinh(\alpha^2) - \alpha^2 - \frac{\alpha^6}{6}\right) \left(\sinh(\beta^2) - \beta^2 - \frac{\beta^6}{6}\right) \nonumber \\
	&+2 \cos\phi \left(\sinh(\alpha \beta) - \alpha \beta - \frac{(\alpha \beta)^3 }{6}\right)^2 \Bigg\},
\end{align}
where the overall factor is given by 
\begin{align}\label{key}
	\Omega_- =  \frac{1}{2} \left[\sinh(\alpha^2) \sinh(\beta^2) + \cos\phi \, \sinh^2(\alpha \beta) \right]^{-1}.
\end{align} 

In the same way as before, we adopt $\phi = \pi$ and the choice of parameters $a = 0,\, a' = \pi/2,\, b = +\pi/4,\, b' = -\pi/4$. In this case, we immediately find for $\langle \mathcal{C} \rangle$:
\begin{align}\label{key}
	\langle \mathcal{C}_{CHSH} \rangle &= \frac{1}{3 \kappa_-} \Big[ \left(\sqrt{2} -1\right) \alpha^2 \beta^2 (\alpha^2 - \beta^2)^2  \nonumber \\
	&+ 6 \sinh^2(\alpha  \beta ) -6 \sinh(\alpha^2) \sinh(\beta^2) \nonumber \\
	&- \alpha \beta \sinh(\alpha \beta) \left[12 + 2 \alpha^2 \beta^2 - (\sqrt{3} + \sqrt{6}) (\alpha^2 + \beta^2)\right] \nonumber \\
	&+ \alpha^2 \sinh(\beta^2) \left[6 - (\sqrt{3} + \sqrt{6}) \alpha^2 + \alpha^4\right]  \nonumber \\
	&+ \beta^2 \sinh(\alpha^2) \left[6 - (\sqrt{3} + \sqrt{6}) \beta^2  + \beta^4\right]  \Big],  
\end{align}
where $\kappa_- = \sinh^2(\alpha  \beta) - \sinh(\alpha^2) \sinh(\beta^2)$. Considering $(\alpha, \beta) = (0.1, 0.2)$ we find $\langle \mathcal{C}_{CHSH} \rangle = 2.8280$ and the Tsirelson's bound is saturated already at $(\alpha, \beta) = (0.08, 0.09)$. Furthermore, here we observe a violation for the whole range of parameters $\alpha$ and $\beta$ between $0$ and $1$. Adopting the parameters $(\alpha, \beta) = (0.1, 0.8)$, we still find violation for relative phases around $\phi \in (\pi - 0.2, \pi + 0.2)$.

\begin{figure}[t!]
	\begin{minipage}[b]{1.0\linewidth}
		\includegraphics[width=\textwidth]{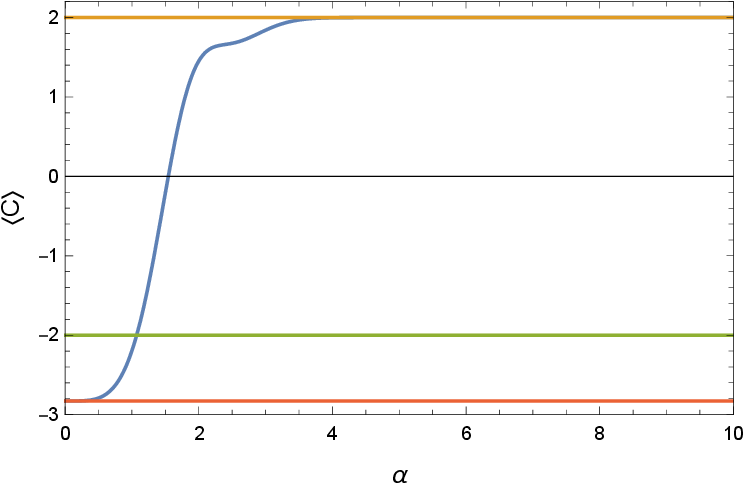}
	\end{minipage} \hfill
	\caption{$\langle \mathcal{C} \rangle$ correlator in blue, as a function of $\alpha$ in the odd cat state, with $\beta = \alpha + 0.001$. Here we considered $\phi = \pi, \, a = 0,\, a' = \pi/2,\, b = +\pi/4,\, b' = -\pi/4$. The red line represents the Tsirelson's bound.}
	\label{BellCatMinus2D}
\end{figure}

\begin{figure}[t!]
	\begin{minipage}[b]{1.0\linewidth}
		\includegraphics[width=\textwidth]{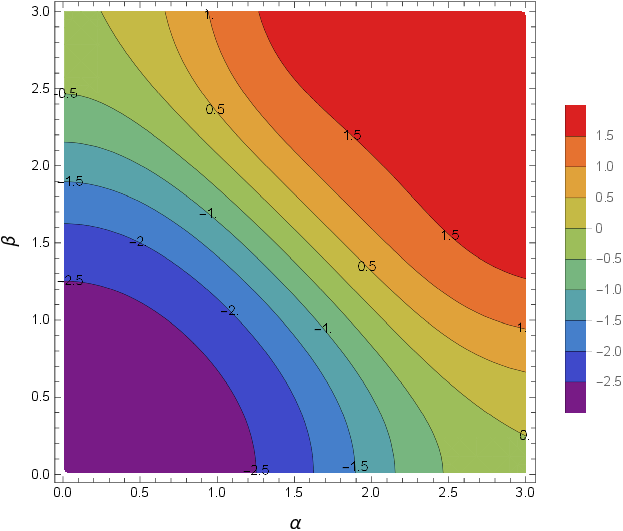}
	\end{minipage} \hfill
	\caption{Contour plot exhibiting $\vert \langle \mathcal{C} \rangle \vert$ as a function of $\alpha$ and $\beta$ in the odd cat state. Here we have phase $\phi = \pi$, and parameters $a = 0,\, a' = \pi/2,\, b = +\pi/4,\, b' = -\pi/4$.}
	\label{BellCatMinusContour}
\end{figure}

The results for the odd cat states are very similar to the ones obtained for even cat states, as one can immediately see in Fig.~\ref{BellCatMinus2D}, where we show $\langle \mathcal{C} \rangle$ as a function of $\alpha$ with $\beta = \alpha + 0.001$, and also in Fig.~\ref{BellCatMinusContour}, where the range of parameters $(\alpha, \beta)$ leading to a violation is shown.

\section{Conclusions}\label{Conclusions}

In this work, we investigated the violation of Bell-CHSH inequalities considering some interesting examples of entangled coherent states. The Bell's operators construction by using the pseudospin operators plays a prominent role, allowing a detailed analysis of the entangled coherent states considered here. We have studied in two different Bell setups three types of entangled coherent states: the symmetric, the asymmetric, and the cat states. In each of them, we computed the Bell-CHSH correlator and presented the set of parameters that leads to the explicit violation of Bell-CHSH inequalities. Furthermore, we highlighted the particular values of $\alpha$ and $\beta$ leading to the saturation of Tsirelson's bound for each case, collecting them in Table~\ref{tabela2}.

It is worth pointing out that in the symmetric case, we need to choose $\alpha$ and $\beta$ small in order to find a violation, and very small to saturate the Tsirelson's bound. On the other hand, in the asymmetric case, we can find a Bell-CHSH inequality violation even with $\alpha$ and $\beta$ near to $1$, but it is important to keep them close to each other, finding a larger violation if we consider the $\alpha=\beta$ situation. Finally, the entangled cat states exhibit Bell-CHSH inequality violation for almost all the values of $\alpha$ and $\beta$ in the interval $(0,1)$. 
In all the above statements, we are tacitly considering a relative phase $\phi = \pi$ and the set of parameters $a = 0, a' = \pi/2, b = +\pi/4, b' = -\pi/4$, but the computations were done for a general set of parameters.
In almost all the cases considered here, one can see that the Bell-CHSH inequality violation occurs essentially for rather small values of the parameters $\alpha, \beta$. As these parameters increase, the Bell-CHSH correlator $\vert \langle \mathcal{C} \rangle \vert$  asymptotes to $2$, a feature that is already clearly visible around $\alpha \approx 3$. We are then led to interpret this pattern in terms of the expected classical behavior: for large values of $\alpha$, the states enter in a quasiclassical regime and the Bell-CHSH inequality violation disappears.

The asymmetric case in the second Bell setup studied in Sec.~\ref{Asymm2} stands as an exception, exhibiting Bell-CHSH inequality violation in the case $\phi=\pi$ for all the considered values of $\alpha$. Furthermore, in this case we were able to find a Bell-CHSH violation even in the situation without a relative phase, something that we didn't observe in the other cases. However, we emphasize that these particular features of asymmetric coherent states~\eqref{AsymmCoherent} were also observed in~\cite{Park15}, although using a different Bell setup.

It would be extremely interesting if one could devise a realistic experimental setup corresponding to the theoretical framework presented here, in order to measure the Bell-CHSH inequality violation for these entangled coherent states and confirm our predictions. Furthermore, the extension of this investigation to the realm of multipartite systems could bring quite interesting developments.  

\vspace{-0.3cm}

\begin{table}[h!]
	\centering
	\begin{tabular}{|c|c|}
		\hline
		State & $\left(\alpha, \beta\right)$ such that $\vert \langle \mathcal{C} \rangle \vert = 2.8284 $
		\\ \hline
		Symm. Coherent~\ref{Symm1}  & $\left(0.001, 0.002\right)$
		\\ \hline 
		Symm. Coherent~\ref{Symm2} & $\left(0.001, 0.002\right)$ 
		\\ \hline 
		Asymm. Coherent~\ref{Asymm1} & $\left(0.06, 0.06\right)$
		\\ \hline 
		Asymm. Coherent~\ref{Asymm2} & $\left(0.1, 0.1\right)$
		\\ \hline 
		Cat Coherent $(+)$~\ref{SecCatCohe} & $\left(0.07, 0.08\right)$
		\\ \hline 
		Cat Coherent $(-)$~\ref{SecCatCohe} & $\left(0.08, 0.09\right)$
		\\ \hline 
	\end{tabular}
	\caption{Parameters $\left(\alpha, \beta\right)$ such that the absolute value of $\langle \mathcal{C}_{CHSH} \rangle$ saturates the Tsirelson's bound $2\sqrt{2} \approx 2.8284$. Here we adopted the relative phase $\phi = \pi$ and the same parameters used in the main text: $a = 0, a' = \pi/2, b = +\pi/4, b' = -\pi/4$.}
	\label{tabela2}
\end{table}

\vspace{-1.3cm}

\begin{acknowledgments}
\vspace{-0.5cm}
The authors thank the Brazilian agencies CNPq, CAPES and FAPERJ for financial support.  S.~P.~Sorella is a level $1$ CNPq researcher under the contract 301030/2019-7. G.~Peruzzo is a FAPERJ postdoctoral fellow in the {\it Pós-Doutorado Nota 10} program under the contracts E-26/205.924/2022 and E-26/205.925/2022. PDF is grateful to Fernando de Melo for the very interesting discussion and also for the useful suggestions.
\end{acknowledgments}


\begin{thebibliography}{99}

\bibitem{EPR}
A. Einstein, B. Podolsky, and N. Rosen, {\it Can Quantum-Mechanical Description of Physical Reality be Considered Complete?}, Physical Review. 47 (10): 777 (1935).


\bibitem{Horodecki09}
R. Horodecki, P. Horodecki, M. Horodecki, and K. Horodecki, {\it Quantum Entanglement}, Rev. Mod. Phys. {\bf 81}, 865 (2009).

\bibitem{Bell64}
J. S. Bell, {\it On the Einstein Podolsky Rosen paradox}, Physics Physique Fizika {\bf 1}, 195 (1964).

\bibitem{Bell66} 
J. S. Bell, {\it On the Problem of Hidden Variables in Quantum Mechanics}, Rev. Mod. Phys. {\bf 38}, 447 (1966).

\bibitem{CHSH69}
J. F. Clauser, M. A. Horne, A. Shimony, and R. A. Holt, {\it Proposed Experiment to Test Local Hidden-Variable Theories}, Phys. Rev. Lett. {\bf 23}, 880 (1969).

\bibitem{CHSH74}
J. F. Clauser and M. A. Horne, {\it Experimental consequences of objective local theories}, Phys. Rev. D {\bf 10}, 526 (1974).

\bibitem{CHSH78}
J. F. Clauser and A. Shimony, {\it Bell’s theorem: Experimental tests and implications}, Rept. Prog. Phys. {\bf 41}, 1881 (1978).



\bibitem{ExpCHSH1}
S. J. Freedman and J. F. Clauser, {\it Experimental test of local hidden-variable theories}, Phys. Rev. Lett. {\bf 28}, 938 (1972).

\bibitem{ExpCHSH2}
A. Aspect, {\it Proposed experiment to test the nonseparability of quantum mechanics}, Phys. Rev. D {\bf 14}, 1944 (1976).

\bibitem{ExpCHSH3}
A. Aspect, P. Grangier, and G. Roger, {\it Experimental Tests of Realistic Local Theories via Bell’s Theorem}, Phys. Rev. Lett. {\bf 47}, 460 (1981).

\bibitem{ExpCHSH4}
A. Aspect, P. Grangier, and G. Roger, {\it Experimental Realization of Einstein- Podolsky-Rosen-Bohm Gedankenexperiment: A New Violation of Bell’s Inequalities}, Phys. Rev. Lett. {\bf 49}, 91 (1982).

\bibitem{ExpCHSH5}
A. Aspect, J. Dalibard, and G. Roger, {\it Experimental test of Bell’s inequalities using time-varying analyzers}, Phys. Rev. Lett. {\bf 49}, 1804 (1982).

\bibitem{ExpCHSH6}
M.A. Rowe et al., {\it Experimental violation of a Bell’s inequality with efficient detection}, Nature {\bf 409}, 791 (2001).

\bibitem{ExpCHSH7}
M. Ansmann et al., {\it Violation of Bell’s inequality in Josephson phase qubits}, Nature {\bf 461}, 504 (2009).

\bibitem{ExpCHSH8}
M. Giustina et al., {\it Bell violation using entangled photons without the fair-sampling assumption}, Nature {\bf 497}, 227 (2013).

\bibitem{ExpCHSH9}
M. Giustina et al., {\it Significant-Loophole-Free Test of Bell’s Theorem with Entangled Photons}, Phys. Rev. Lett. {\bf 115}, 250401 (2015).

\bibitem{Tsirelson80}
B. S. Cirel’son, {\it Quantum generalizations of Bell’s inequality}, Lett. Math. Phys. {\bf 4}, 93 (1980). 

\bibitem{Tsirelson87}
B. S. Tsirel’son, {\it Quantum analogues of the Bell inequalities. The case of two spatially separated domains}, J. Math. Sci. {\bf 36}, 557 (1987).




\bibitem{Werner01}
R. F. Werner and M. M. Wolf, {\it Bell inequalities and entanglement}, Quantum Inf. Comput. {\bf 1}, 1 (2001).

\bibitem{Yu21}
Y. Yu, {\it Advancements in Applications of Quantum Entanglement}, J. Phys.: Conf. Ser. {\bf 2012}, 012113 (2021).



\bibitem{Schrodinger26}
E. Schr\"odinger, {\it Der stetige Übergang von der Mikro- zur Makromechanik}, Naturwissenschaften {\bf 14}, 664 (1926).

\bibitem{Klauder85}
J. R. Klauder and B. S. Skagerstam, Coherent States	(World Scientific, Singapore, 1985).

\bibitem{Glauber63}
R. J. Glauber, {\it Photon Correlations}, Phys. Rev. Lett. {\bf 10}, 84 (1963).

\bibitem{Sudarshan63}
E. C. G. Sudarshan, {\it Equivalence of Semiclassical and Quantum Mechanical Descriptions of Statistical Light Beams}, Phys. Rev. Lett. {\bf 10}, 277 (1963).

\bibitem{SandersReview}
B. C. Sanders, {\it Review of entangled coherent states}, J. Phys. A: Math. Theor. {\bf 45}, 244002 (2012).

\bibitem{Milburn85}
G. J. Milburn, {\it Quantum and classical Liouville dynamics of the anharmonic oscillator}, Phys. Rev. A {\bf 33}, 674 (1986).

\bibitem{Milburn86}
G. J. Milburn and C. A. Holmes, {\it Dissipative Quantum and Classical Liouville Mechanics of the Anharmonic Oscillator}, Phys. Rev. Lett. {\bf 56}, 2237 (1986).

\bibitem{Yurke86}
B. Yurke and D. Stoler, {\it Generating quantum mechanical superpositions of macroscopically distinguishable states via amplitude dispersion}, Phys. Rev. Lett. {\bf 57}, 13 (1986).

\bibitem{Yurke87}
B. Yurke and D. Stoler, {\it Quantum behavior of a four-wave mixer operated in a nonlinear regime}, Phys. Rev. A {\bf 35}, 4846 (1987).

\bibitem{Glancy08}
S. Glancy and H. M. de Vasconcelos, {\it Methods for producing optical coherent state superpositions}, J. Opt. Soc. Am. B {\bf 25}, 712 (2008).

\bibitem{Titulaer65}
U. M. Titulaer and R. J. Glauber, {\it Density Operators for Coherent Fields}, Phys. Rev. {\bf 145}, 1041 (1965).

\bibitem{Birula68}
Z. Bialynicka—Birula, {\it Properties of the generalized coherent state}, Phys. Rev. {\bf 173}, 1207 (1968).

\bibitem{Stoler71}
D. Stoler, {\it Generalized Coherent States}, Phys. Rev. D {\bf 4}, 2309 (1971).

\bibitem{Arecchi72}
F. T. Arecchi, E. Courtens, R. Gilmore, and H. Thomas, {\it Atomic Coherent States in Quantum Optics}, Phys. Rev. A {\bf 6}, 2211 (1972).

\bibitem{Gilmore72}
R. Gilmore, {\it Geometry of symmetrized states}, Ann. Phys. {\bf 74}, 391 (1972).

\bibitem{Perelomov72}
A. M. Perelomov, {\it Coherent states for arbitrary Lie group}, Commun. Math. Phys. {\bf 26}, 222 (1972).

\bibitem{Perelomov86}
A. M. Perelomov, {\it Generalized Coherent States and Their Applications}, Springer Berlin (1986).

\bibitem{Barut71}
A. O. Barut and L. Girardello, {\it New “coherent” states associated with non-compact groups}, Commun. Math. Phys. {\bf 21}, 41 (1971).

\bibitem{Buzek95}
V. Bu\v{z}ek and P. L. Knight, {\it I: Quantum Interference, Superposition States of Light, and Nonclassical Effects}, Prog. Opt. {\bf 34}, 1 (1995).

\bibitem{Aharonov67}
Y. Aharonov and L. Susskind, {\it Charge superselection rule} Phys. Rev. {\bf 155}, 1428 (1967).

\bibitem{Agarwal86}
G. S. Agarwal, {\it Generation of pair coherent states and squeezing via the competition of four-wave mixing and amplified spontaneous emission}, Phys. Rev. Lett. {\bf 57}, 827 (1986).

\bibitem{Agarwal88}
G. S. Agarwal, {\it Nonclassical statistics of fields in pair coherent states},  J. Opt. Soc. Am. B {\bf 5}, 1940 (1988).

\bibitem{Agarwal91}
G. S. Agarwal and K. Tara, {\it Nonclassical properties of states generated by the excitations on a coherent state}, Phys. Rev. A {\bf 43}, 492 (1991).


\bibitem{Mecozzi87}
A. Mecozzi and P. Tombesi, {\it Distinguishable quantum states generated via nonlinear birefringence},  Phys. Rev. Lett. {\bf 58}, 1055 (1987).

\bibitem{Tombesi87}
A. Mecozzi and P. Tombesi, {\it Generation of macroscopically distinguishable quantum states and detection by the squeezed-vacuum technique}, J. Opt. Soc. Am. B {\bf 4}, 1700 (1987).

\bibitem{Sanders92}
B. C. Sanders, {\it Entangled coherent states}, Phys. Rev. A {\bf 45}, 6811 (1992); {\it Erratum}, Phys. Rev. A {\bf 46}, 2966 (1992).



\bibitem{Mann95}
A. Mann, B. C. Sanders, and W. J. Munro, {\it Bell's inequality for an entanglement of nonorthogonal states}, Phys. Rev. A {\bf 51}, 989 (1995).

\bibitem{Jex95}
I. Jex, P. T\"orm\"a, and S. Stenholm, {\it Multimode coherent states}, J. Mod. Opt. {\bf 42}, 1377 (1995).

\bibitem{Zheng98}
S-B. Zheng, {\it A scheme for the generation of multi-mode Schrödinger cat states}, Quantum. Semiclass. Opt. {\bf 10}, 691 (1998).

\bibitem{Wang01b}
X. Wang and B. C. Sanders, {\it Multipartite entangled coherent states}, Phys. Rev. A {\bf 65}, 012303 (2001).

\bibitem{Jeong06}
H. Jeong and N. B. An, {\it Greenberger-Horne-Zeilinger–type and W-type entangled coherent states: Generation and Bell-type inequality tests without photon counting}, Phys. Rev. A {\bf 74}, 022104 (2006).

\bibitem{Li09}
H-M. Li, H-C. Yuan, and H-Y. Fan, {Single-mode excited GHZ-type entangled coherent state}, Int. J. Theor. Phys. {\bf 48}, 2849 (2009).

\bibitem{Munhoz08}
P. P. Munhoz, F. L. Semião, A. Vidiella-Barranco, and J. A. Roversi, {\it Cluster-type entangled coherent states}, Phys. Lett. A {\bf 372}, 3580 (2008).

\bibitem{Becerra08}
E. M. Becerra-Castro, W. B. Cardoso, A. T. Avelar, and B. Baseia, {\it Generation of a 4-qubit cluster of entangled coherent states in bimodal QED cavities}, J. Phys. B: At. Mol. Opt. Phys. {\bf 41}, 085505 (2008).

\bibitem{Wang08}
W. Wen-Feng et al., {\it Generation of Cluster-Type Entangled Coherent States via Cross-Kerr Nonlinearity}, Chinese Phys. Lett. {\bf 25} 839 (2008).

\bibitem{Wang01a}
X. Wang, {\it Quantum teleportation of entangled coherent states}, Phys. Rev. A {\bf 64}, 022302 (2001).

\bibitem{Bennett93}
C. H. Bennett, G. Brassard, C. Crépeau, R. Jozsa, A. Peres, and W. K. Wootters, {\it Teleporting an unknown quantum state via dual classical and Einstein-Podolsky-Rosen channels}, Phys. Rev. Lett. {\bf 70}, 1895 (1993).



\bibitem{Janszky02}
J. Janszky, A. Gabris, M. Koniorczyk, A. Vukics, and J. K. Asb´oth, {\it One-complex-plane representation: a coherent-state description of entanglement and teleportation}, J. Opt. B: Quantum Semiclass. Opt. 4 S213 (2002).

\bibitem{Johnson02}
T. J. Johnson, S. D. Bartlett, and B. C. Sanders, {\it Continuous-variable quantum teleportation of entanglement}, Phys. Rev. A {\bf 66}, 042326 (2002).

\bibitem{Jeong01}
H. Jeong, M. S. Kim, and J. Lee, {\it Quantum-information processing for a coherent superposition state via a mixedentangled coherent channel}, Phys. Rev. A {\bf 64}, 052308 (2001).

\bibitem{vanEnk01}
S. J. van Enk and O. Hirota, {\it Entangled coherent states: Teleportation and decoherence}, Phys. Rev. A {\bf 64}, 022313 (2001).

\bibitem{Munro01}
W. J. Munro, G. J. Milburn, and B. C. Sanders, {\it Entangled coherent-state qubits in an ion trap}, Phys. Rev. A {\bf 62}, 052108 (2001).

\bibitem{Cochrane99}
P. T. Cochrane, G. J. Milburn, and W. J. Munro, {\it Macroscopically distinct quantum-superposition states as a bosonic code for amplitude damping}, Phys. Rev. A {\bf 59}, 2631 (1999).

\bibitem{Oliveira00}
M. C. de Oliveira and W. J. Munro, {\it Quantum computation with mesoscopic superposition states}, Phys. Rev. A {\bf 61}, 042309 (2000).

\bibitem{vanLoock08}
P. van Loock, N. L\"utkenhaus, W. J. Munro, and K. Nemoto, {\it Quantum repeaters using coherent-state communication}, Phys. Rev. A {\bf 78}, 062319 (2008).

\bibitem{El Allati11}
A. El Allati, Y. Hassouni, and N. Metwally, {\it Communication via an entangled coherent quantum network}, Phys. Scr. {\bf 83}, 065002 (2011).

\bibitem{Ansari94a}
N. A. Ansari et al., {\it Quantum limits in interferometric gravitational-wave antennas in the presence of even and odd coherent states}, Phys. Rev. A {\bf 49}, 2151 (1994).

\bibitem{Ansari94b}
N. A. Ansari and V. I. Man’ko, {\it Photon statistics of multimode even and odd coherent light}, Phys. Rev. A {\bf 50}, 1942 (1994).

\bibitem{Gerry01}
C. C. Gerry and R. A. Campos, {\it Generation of maximally entangled photonic states with a quantum-optical Fredkin gate}, Phys. Rev. A {\bf 64}, 063814 (2001).

\bibitem{Gerry02}
C. C. Gerry, A. Benmoussa, and R. A. Campos, {\it Nonlinear interferometer as a resource for maximally entangled photonic states: Application to interferometry}, Phys. Rev. A {\bf 66}, 013804 (2002).

\bibitem{Joo11}
J. Joo, W. J. Munro, and T. P. Spiller, {\it Quantum Metrology with Entangled Coherent States}, Phys. Rev. Lett. {\bf 107}, 083601 (2011); {\it Erratum}, Phys. Rev. Lett. 107, 219902 (2011).


\bibitem{psi1}
Z. B. Chen, J. W. Pan, G. Hou and Y. D. Zhang, {\it Maximal violation of Bell's inequalities for continuous variable systems}, Phys. Rev. Lett. {\bf 88}, 040406 (2002). 

\bibitem{psi2}
J. A. Larsson, {\it Qubits from number states and Bell inequalities for number measurements}, Phys. Rev. A {\bf 67}, 022108 (2003).

\bibitem{psi3} 
M. M. Dorantes and J. L. Lucio M., {\it Generalizations of the pseudospin operator to test the Bell inequality for the TMSV state}, J. Phys. A: Math. Theor. {\bf 42}, 285309 (2009). 




\bibitem{Dowling08}
J. P. Dowling, {\it Quantum optical metrology–the lowdown on high-N00N states}, Contemp. Phys. {\bf 49}, 125 (2008).

\bibitem{Boto00}
A. N. Boto et al., {\it Quantum Interferometric Optical Lithography: Exploiting Entanglement to Beat the Diffraction Limit}, Phys. Rev. Lett. {\bf 85}, 2733 (2000).

\bibitem{Wildfeuer07}
C. F. Wildfeuer, A. P. Lund, and J. P. Dowling, {\it Strong violations of Bell-type inequalities for path-entangled number states}, Phys. Rev. A {\bf 76}, 052101 (2007).

\bibitem{Banaszek99}
K. Banaszek and K. Wodkiewicz, {\it Testing quantum nonlocality in phase space}, Phys. Rev. Lett. {\bf 82}, 2009 (1999).

\bibitem{Gerry09}
C. C. Gerry, J. Mimih, and A. Benmoussa, {\it Maximally entangled coherent states and strong violations of Bell-type inequalities}, Phys. Rev. A {\bf 80}, 022111 (2009).

\bibitem{Gerry00}
C. C. Gerry, {\it Heisenberg-limit interferometry with four-wave mixers operating in a nonlinear regime}, Phys. Rev. A {\bf 61}, 043811 (2000).

\bibitem{Benmoussa02}
C. C. Gerry and A. Benmoussa, {\it Heisenberg-limited interferometry and photolithography with nonlinear four-wave mixing}, Phys. Rev. A {\bf 65}, 033822 (2002).


\bibitem{Park15}
C-Y. Park and H. Jeong, {Bell-inequality tests using asymmetric entangled coherent states in asymmetric lossy environments}, Phys. Rev. A {\bf 91}, 042328 (2015).


\bibitem{Dodonov74}
V. V. Dodonov, I. A. Malkin, and V. I. Man'ko, {\it Even and odd coherent states and excitations of a singular oscillator}, Physica {\bf 72}, 597 (1974).


\bibitem{Gisin} 
N. Gisin and A. Peres, {\it Maximal violation of Bell's inequality for arbitrarily large spin}, Phys. Lett A {\bf 162}, 15 (1992).

\bibitem{Peruzzo:2023nrr}
G. Peruzzo and S. P. Sorella, {\it Entanglement and maximal violation of the CHSH inequality in a system of two spins j: A novel construction and further observations}, Phys. Lett. A \textbf{474}, 128847 (2023).

\bibitem{Sorella:2023hku}
S. P. Sorella, {\it A study of the violation of the Bell-CHSH inequality}, arXiv:2302.02385.

\bibitem{Sorella:2023iwz}
S. P. Sorella, {\it On the representations of Bell's operators in Quantum Mechanics}, Found. Phys. {\bf 53}, 59 (2023).






















\end{thebibliography}
\end{document}